\documentclass[twocolumn,bibnotes]{revtex4}

\usepackage{color,rotating,graphicx}
\usepackage{amsmath,ulem, comment}
\usepackage{dsfont}

\renewcommand{\Im}{{\rm Im}}
\newcommand{\ri}{{\rm i}}
\newcommand{\re}{{\rm e}}
\newcommand{\rd}{{\rm d}}

\newcommand{\kb}{k_{\rm B}}

\begin{document}

\author{S.-A. Biehs}
\email{s.age.biehs@uol.de}

\affiliation{Institut f\"{u}r Physik, Carl von Ossietzky Universit\"{a}t, D-26111 Oldenburg, Germany}
\affiliation{Center for Nanoscale Dynamics  (CeNaD), Carl von Ossietzky Universit\"{a}t, D-26129 Oldenburg, Germany}
\affiliation{Laboratoire Charles Coulomb (L2C), UMR 5221 CNRS-University of Montpellier, F-34095 Montpellier, France}

\author{P. Rodriguez-Lopez}
\email{pablo.ropez@urjc.es}
\affiliation{\'{A}rea de Electromagnetismo and Grupo Interdisciplinar de Sistemas Complejos (GISC), Universidad Rey Juan Carlos, 28933 M\'{o}stoles, Madrid, Spain}
\affiliation{Laboratoire Charles Coulomb (L2C), UMR 5221 CNRS-University of Montpellier, F-34095 Montpellier, France}

\author{M. Antezza}
\email{mauro.antezza@umontpellier.fr}
\affiliation{Laboratoire Charles Coulomb (L2C), UMR 5221 CNRS-University of Montpellier, F-34095 Montpellier, France}
\affiliation{Institut Universitaire de France, 1 rue Descartes, Paris Cedex 05, F-75231, France}

\author{G. S. Agarwal}
\email{girish.agarwal@tamu.edu}
\affiliation{ Institute for Quantum Science and Engineering and Department of Biological and 
Agricultural Engineering Department of Physics and Astronomy, Texas A \& M University, College Station, Texas 77845, USA}

\title{Nonreciprocal heat flux via synthetic fields in linear quantum systems}

\begin{abstract}
	We study the heat transfer between N coupled quantum resonators with applied synthetic electric and magnetic fields realized by changing the resonators parameters by external drivings. To this end we develop two general methods, based  on the quantum optical master equation and on the Langevin equation for $N$ coupled oscillators where all quantum oscillators can have their own heat baths. The synthetic electric and magnetic fields are generated by a dynamical modulation of the oscillator resonance with a given phase. Using Floquet theory we solve the dynamical equations with both methods which allow us to determine the heat flux spectra and the transferred power. With apply these methods to study the specific case of a linear tight-binding chain of four quantum coupled resonators. We find that in that case, in addition to a non-reciprocal heat flux spectrum already predicted in previous investigations, the synthetic fields induce here non-reciprocity in the total heat flux hence realizing a net heat flux rectification.

\end{abstract}

\maketitle

\section{Introduction}

In the last decade a great number of experiments have verified the near-field enhancement of thermal radiation between two macroscopic objects down to distances of a few nanometer~\cite{HuEtAl2008,OttensEtAl2011,Kralik2012,LimEtAl2015,WatjenEtAl2016,BernadiEtAl2016,SongEtAl2016,FiorinoEtAl2018b,Mittapally2023}. In particular, the theoretically proposed effects of thermal rectification with a phase-change diode~\cite{Ito,FiorinoEtAl2018a}, a phase-change material based memory~\cite{ItoMem} and active heat flux switching or modulations~\cite{Minnich,Minnich2,ShiEtAl2022} have been realized, experimentally. Also several proposals for non-reciprocal heat transport have been made, but these effects have not been demonstrated experimentally, yet. Typically, these proposals rely on the application of magnetic fields to nanoscale setups involving magneto-optical materials or by using Weyl semi-metals with intrinsic nonreciprocal optical properties. It can be shown theoretically that by means of magnetic fields the magnitude of the heat flux and its direction can be manipulated~\cite{Latella2017,He2020,Cuevas,reviewMagneto,MoncadaVilla1,MoncadaVilla2,GeEtAl2019,WuEtAl2021}. Due to the broken time-reversal symmetry also non-reciprocal heat fluxes can exist in such cases leading to persistent heat currents and fluxes~\cite{zhufan,CircularHeatFlux}, persistent angular momenta and spins~\cite{Silveirinha,CircularHeatFlux,Zubin2019}, normal and anomalous Hall effect for thermal radiation~\cite{hall,ahall}, and diode effects by coupling to non-reciprocal surface modes~\cite{Doyeux2017,NRdiode,HuEtAl2023}, and spin-directional near- and far-field thermal emission~\cite{FanEtAl2020,DongEtAl2021}. A tradeoff of using magneto-optical materials is that for having observable non-reciprocal heat fluxes, experiments with large magnetic fields in a nanoscale setup are necessary. On the other hand, using Weyl semi-metals with intrinsic nonreciprocity does not allow for a dynamic tuning. 

\begin{figure}
  \includegraphics[angle=0,scale=0.35]{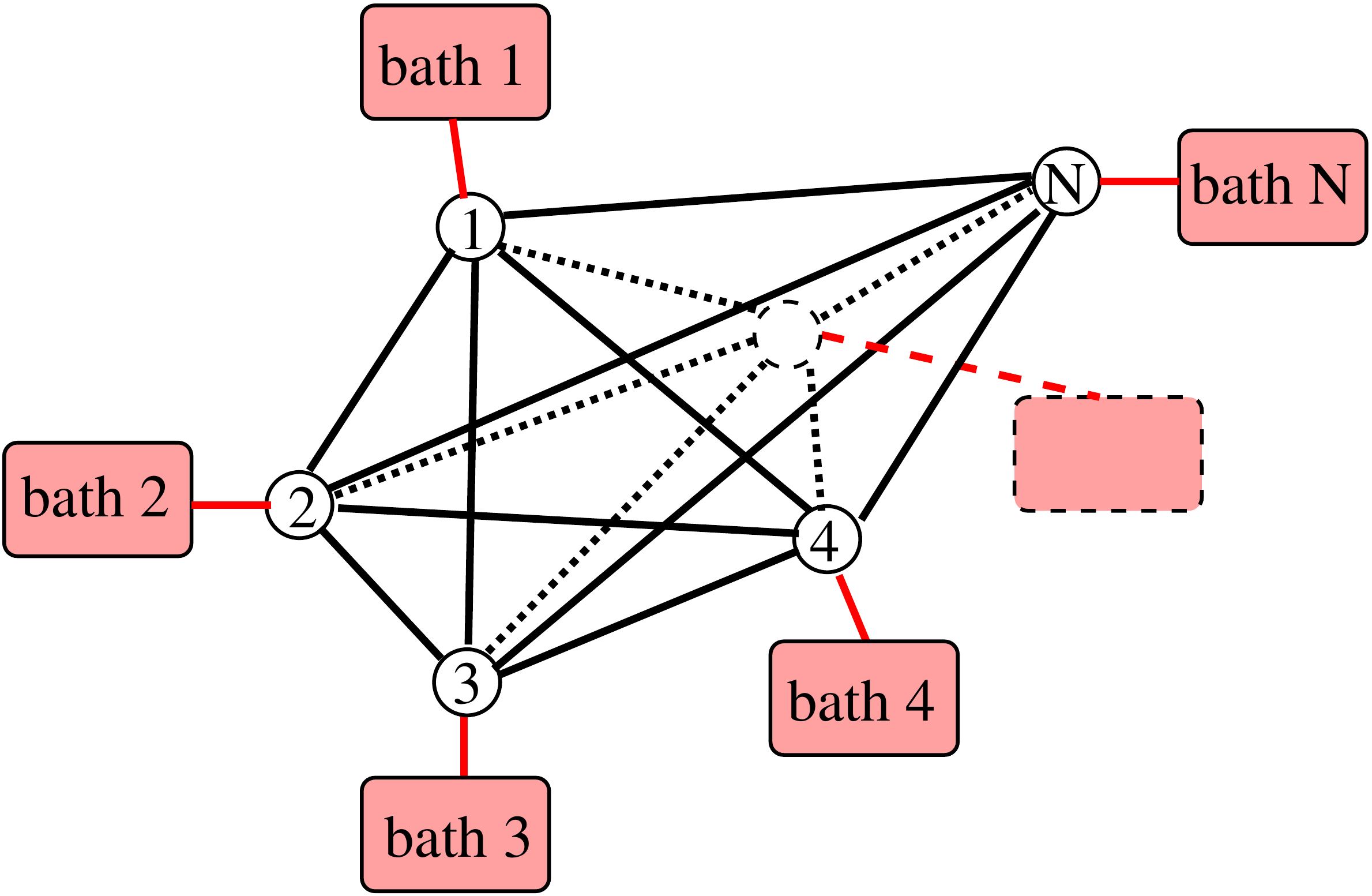}
	\caption{\label{Fig:NOscillators} Sketch of $N$ coupled quantum resonators each coupled to its own heat bath.}
\end{figure}

Recently, the modulation of resonance frequencies of a system of resonators with a single modulation frequency but different phases has been interpreted as a mean to create synthetic electric and magnetic fields~\cite{Lustig}. For the energy transmission in a setup of two resonators with applied synthetic electric and magnetic fields, i.e.\ with a modulation of the resonance frequencies and a phase shift, it could be shown experimentally and theoretically that monochromatic waves are transmitted in a nonreciprocal manner~\cite{PetersonEtAl2019} if there is a non-zero phase shift, i.e.\ a synthetic magnetic field. Now, if the two resonators with applied synthetic electric and magnetic fields are coupled to two thermal reservoirs within a master equation approach~\cite{JOSA,Barton,Karthik,Antezza} then the transmission coefficients for the heat current in both directions are not the same which is a manifestation of a broken detailed balance~\cite{SABGSA2023}. However, in this case the total transferred power between both resonances itself is reciprocal even in presence of synthetic electric and magnetic fields~\cite{SABGSA2023}. 

That the transferred power is reciprocal might not be surprising for two reasons. First of all, in the context of Rytov's fluctuational electrodynamics it can easily shown that the total radiative heat flux between two objects is always reciprocal~\cite{Latella2017}. Non-reciprocal effects necessitate at least a third object and non-reciprocal material properties of the objects or environment~\cite{Herz,RMP}. Another argument is that within the quantum master equation approach for linearly coupled oscillators typically non-linear effects need to be included to have nonreciprocal heat flow~\cite{LandiEtAl2022} even though it seems that nonreciprocal heat flow can also be generated by specific choices of temperatures in a linear chain of oscillators~\cite{PurkayasthaEtAl2016,KalantarEtAl2021}. However, as we will show below the application of synthetic electric and magnetic fields can indeed generate nonreciprocal heat flow in a tight-binding configuration of four coupled resonators without the need of non-linearity due to the presence of the synthetic magnetic field.

Now, we distinguish our work from previous studies. Several kinds of modulations have been proposed like the periodic modulation of the permittivity~\cite{LiraEtAl2012,ChamanaraEtAl2017,EstepEtAl2014}. Such modulations have been shown to introduce synthetic magnetic fields for photons~\cite{TzuanEtAl2014} and consequently related effects like the Aharonov-Bohm effect for photons~\cite{FangEtAl2012}. In the context of thermal radiation it could be demonstrated that permittivity modulations can introduce nonreciprocity which manifests itself in a breakdown of the detailed balance in Kirchhoff's law~\cite{GhanekarEtAl2022} and can be employed for photonic refrigeration~\cite{BuddhirajuEtAl2020}. In similar approaches a combined dynamical modulation of the resonances of heat exchanging objects and their interaction strength were applied resulting in a heat pumping effect and non-reciprocal heat fluxes in a three resonator configurations~\cite{LiEtAl2019,AlcazarEtAl2021}. Heat pumping effects also exist when only the interaction strengths in three-body configurations are dynamically modulated~\cite{MessinaPBA2020}. It must be emphasized that these effects are different from the heat shuttling effect where temperatures or chemical potential of two reservoirs are periodically modulated around their equilibrium values in order to have a heat transport despite the fact that the system is in average in equilibrium~\cite{LiEtAl2008, LiEtAl2009,Latella}. Indeed, in that case the modulation affects the baths only and not resonator parameters. Finally, it could be demonstrated theoretically that adiabatic dynamical modulation of resonators with non-reciprocal conductances geometrical phases can increase or reduce the thermal relaxation~\cite{SABPBABerry} and rapid magnetic field modulations in magneto-optical systems can substantially increase the cooling~\cite{Messina}. 

In this work, we extend the quantum Langevin equation (qLE) and quantum master equation (qME) approach used in Ref.~\cite{SABGSA2023} to the case of $N$ coupled arbitrary resonators with their own heat baths as sketched in Fig.~\ref{Fig:NOscillators} with applied syntetic electric and magnetic synthetic fields. Both methods can be used to calculate the heat flux between any two resonators which are coupled to their own reservoirs. We show numerically that both methods give the same values for the heat flux. The qLE approach naturally allows for calculating the heat flux spectra, whereas the master-equation method is a better choice for fast numerical calculations of the heat flux. We use both methods to show that the heat flux itself is nonreciprocal in the presence of synthetic fields in a linear tight-binding chain of four-resonators.
 
The manuscript is organized as follows: First, in Sec.~II we introduce the standard master-equation for $N$ coupled resonators with $N$ reservoirs. We derive the dynamical equations for the mean values of products of the resonator amplitudes and introduce the qLE for the coupled resonator system. In Sec.~III we introduce the synthetic fields in the qLE approach and provide a formal solution in Fourier space. In Sec.~IV we introduce the synthetic fields in the master-equation approach and give a formal solution by making a Fourier series ansatz. In Sec.~V we show the occurrence of nonreciprocal heat flux in presence of synthetic electric and magnetic fields in a four-resonator chain. We conclude with a summary of the main results in Sec.~VI.

\section{Langevin and master equations}

We start with writing down the Hamiltonian of a coupled harmonic oscillator system (each oscillator coupled to its own heat bath of oscillators) which is given by~\cite{AgarwalBook2012}
\begin{equation}
  H = H_S + \sum_i H_{B,i} + \sum_i H_{SB,i}
\end{equation}
with the Hamiltonian of the system of coupled oscillators
\begin{equation}
	H_S = \sum_i \hbar \omega_i a^\dagger_i a_i + \sum_{i,j,i\neq j} \hbar g_{ij} a^\dagger_i a_j, 
\end{equation}
with resonance frequencies $\omega_i$ and coupling constants $g_{ij} = g_{ji}^*$ for hermitian system $H_S^\dagger = H_S$ and the bosonic creation and annihilation operators $a^\dagger_i$ and $a_i$. The bath oscillator Hamiltonians are given by($i = 1, \ldots, N$)
\begin{equation}
  H_{B,i} = \sum_j \hbar \omega_{ij} b_{ij}^\dagger b_{ij}
\end{equation}
with bosonic  creation and annihilation operators $b^\dagger_{ij}$ and $b_{ij}$
and the Hamiltonians describing the linear coupling between the system oscillators and their baths are given by
\begin{equation}
  H_{SBi} =  \ri \hbar \sum_j g_{B,ij} (a_i + a^\dagger_i) (b_{ij} - b_{ij}^\dagger)  
\end{equation}
with corresponding coupling constants $g_{B,ij}$.
By assuming the validity of the Born-Markov approximation and tracing out the bath variables one can arrive at the qME~\cite{AgarwalBook2012}
\begin{equation}
\begin{split}
  \frac{\partial \rho_S}{\partial t} &= -\ri \sum_i \omega_i [a^\dagger_i a_i, \rho_S] \\
	      &\quad - \ri \sum_{i,j; i\neq j} g_{ij} [a^\dagger_i a_j, \rho_S] \\
              &\quad - \sum_i \kappa_i (n_i + 1)\bigl(a^\dagger_i a_i \rho_S - 2 a_i \rho_S a^\dagger_i + \rho_S a^\dagger_i a_i \bigr) \\
              &\quad - \sum_i \kappa_i n_i \bigl( a_i a^\dagger_i \rho_S - 2 a^\dagger_i \rho_S a_i + \rho_S a_i a^\dagger_i \bigr) 
\end{split}
\label{Eq:MasterEq}
\end{equation}
where the coupling to the bath oscillators is formally given in terms of the coupling constants $\kappa_{i} = \pi \sum_j g_{B,ij}^2 \delta (\omega_{ij} - \omega_i)$ and $n_{i} = [\exp(\hbar \omega_i/ \kb T_i) - 1]^{-1}$ are the mean occupation numbers at the bath temperatures $T_i$. For the sake of generality we assume that $g_{ij} \neq g_{ji}$ which will allow to include systems without ``inversion symmetry''.

From the qME we can derive the dynamical equation for the mean values of any observable. For example, for the mean values of products of raising and lowering operators we obtain the set of equations ($k,l = 1, \ldots, N; k \neq l$) 
\begin{align}
	\frac{\rd}{\rd t} \langle a^\dagger_k  a_k \rangle &= - \ri \sum_{j, j \neq k} \bigl( g_{kj} \langle a^\dagger_k a_j \rangle - g_{jk} \langle a_k a^\dagger_j \rangle \bigr) \nonumber \\ &\quad - 2 \kappa_k \langle a^\dagger_k a_k \rangle + 2 \kappa_k n_k, \label{Eq:aa}\\
	\frac{\rd}{\rd t} \langle a^\dagger_k a_l\rangle &= {\Omega}_{kl} \langle a^\dagger_k a_l\rangle - \ri \sum_{j \neq k; j\neq l} \bigl( g_{lj} \langle a^\dagger_k a_j\rangle - g_{jk} \langle a^\dagger_j a_l \rangle  \bigr) \nonumber \\  
	&\quad - \ri g_{lk} \bigl( \langle a_k^\dagger a_k \rangle -\langle a_l^\dagger a_l \rangle \bigr) \label{Eq:ab}
\end{align}
with
\begin{equation}
	{\Omega}_{kl} = \ri (\omega_k - \omega_l) - \kappa_k - \kappa_l. 
\end{equation}
In the following we will refer to this set of equations for the mean values of operator products (\ref{Eq:aa}) and (\ref{Eq:ab}) as master-equation approach as they are derived from the qME in Eq.~(\ref{Eq:MasterEq}). 

Similarly, one obtains for the time evolution of the mean values of the raising/lowering operators of each oscillator $a_i$ the set of equations ($k = 1, \ldots, N$)
\begin{equation}
	\frac{\rd}{\rd t} \langle a_k \rangle = - \Omega_k \langle a_k \rangle - \ri \sum_{i; i \neq k} g_{ki} \langle a_i \rangle  
	  \label{Eq:MeanValue}
\end{equation}
with $\Omega_k \equiv \ri \omega_k + \kappa_k$. The set of equations for the mean values of the lowering operators of the two oscillators in Eq.~(\ref{Eq:MeanValue}) motivates the introduction of a set of qLE for the operators themselves instead for their expectation values
\begin{equation}
	\dot{a}_k = - \ri \omega_k a_k - \kappa_k a_k - \ri \sum_{i, i\neq k} g_{ki} a_i + F_k, 
\end{equation}
where the coupling to baths is taken into account by the bath operators $F_k$ which obviously must fulfill $\langle F_k \rangle = 0$ to retrieve Eqs.~(\ref{Eq:MeanValue}). Furthermore, to be consistent with the set of Eqs.~(\ref{Eq:aa})-(\ref{Eq:ab}) the correlation functions of the bath operators are given by
\begin{align}
	\langle F^\dagger_{k}(t) F_{k}(t')\rangle &= 2 \kappa_{k} n_{k} \delta(t - t'), \\
	\langle F_{k}(t) F^\dagger_{k}(t')\rangle &= 2 \kappa_{k} (n_{k} + 1)\delta(t - t') 
\end{align}
and $\langle F_{k} F_{k} \rangle = \langle F_{k}^\dagger F_{k}^\dagger \rangle = 0$. Furthermore, the bath operators of different baths are uncorrelated.

\section{Langevin equations with synthetic fields}

We now use the set of qLEs as introduced above and include a frequency modulation ($k = 1, \ldots,N$)
\begin{align}
	\omega_{k} &\rightarrow \omega_k + m_k \beta \cos(\Omega t + \theta_k),  \label{Eq:omegai}
\end{align}
with phase shifts $\theta_k$ and $m_k = \{0, 1\}$ ($m_k = 0$ modulation of oscillator $k$ turned of, $m_k = 1$ modulation turned on). The set of coupled qLE in frequency space is therefore ($k = 1, \ldots,N$)
\begin{equation}
	X_k a_k(\omega) + \ri \sum_{l \neq k} g_{kl} a_l(\omega) =  F_k + \frac{\beta}{2 \ri} \bigl( a_{k,-} \re^{- \ri \theta_k} + a_{k,+} \re^{+ \ri \theta_k} \bigr)
	\label{Eq:SyntheticLagnevin}
\end{equation}
introducing 
\begin{equation}
   X_k = \ri (\omega_k - \omega) + \kappa_{k}
\end{equation}
and the short-hand notation
\begin{equation}
	a_{k,\pm} = a_k(\omega \pm \Omega).
\end{equation}

The coupled qLEs can now be brought in matrix form
\begin{equation}
	\boldsymbol{\psi} = \mathds{M} \mathbf{F} + \frac{\beta}{2 \ri} \mathds{M} \mathds{Q}_+ \boldsymbol{\psi}_+ + \frac{\beta}{2 \ri} \mathds{M} \mathds{Q}_- \boldsymbol{\psi}_- 
\label{Eq:coupledLangevin}
\end{equation}
by introducing the vectors 
\begin{equation}
	\boldsymbol{\psi} = \begin{pmatrix} a_1(\omega) \\ \vdots \\ a_N(\omega) \end{pmatrix}, \boldsymbol{\psi}_\pm = \begin{pmatrix} a_1(\omega \pm \Omega) \\ \vdots \\ a_N(\omega \pm \Omega) \end{pmatrix}, \mathbf{F} = \begin{pmatrix} F_1(\omega) \\ \vdots \\ F_N(\omega) \end{pmatrix}, 
\end{equation}
and the matrices
\begin{equation}
	\mathds{M} = \mathds{A}^{-1} \quad \text{with} \quad 
	\mathds{A} = \begin{pmatrix} X_1        & \ri g_{12}  & \ldots & \ri g_{1N} \\ 
	                             \ri g_{21} & X_2         & \ldots & \ri g_{2N} \\  
	                              \vdots    &  \dots      & \vdots & \vdots \\ 
	                             \ri g_{N1} &  g_{N2}     &  \ldots& X_N \end{pmatrix}
\end{equation}
and
\begin{equation}
	\mathds{Q}_\pm = {\rm diag} (\re^{\pm \ri \theta_1} m_1, \ldots, \re^{\pm \ri \theta_N} m_N).
\end{equation}
In Eq.~(\ref{Eq:coupledLangevin}) it can be clearly seen that due to the modulation there are couplings to the next sidebands $\omega \pm \Omega$ so that this set of equations is recursive and infinitely large. These side bands can be understood as being the consequent of a synthetic constant electric field. Furthermore, the phase shift adds a phase $\pm \theta_k$ to this coupling which can be understood as a consequence of a synthetic magnetic field.

The solution of the coupled qLEs~(\ref{Eq:coupledLangevin}) can formally be written down for all orders. By introducing the block vectors
\begin{align}
	\uline{\boldsymbol{\psi}} &= (\ldots, \boldsymbol{\psi}_{++},\boldsymbol{\psi}_+, \boldsymbol{\psi}, \boldsymbol{\psi}_-, \boldsymbol{\psi}_{--} \ldots)^t, \\
	\uline{\mathbf{F}}  &=  (\ldots, \mathbf{F}_{++}, \mathbf{F}_+, \mathbf{F}, \mathbf{F}_-, \mathbf{F}_{--}, \ldots)^t,
\end{align}
the diagonal block matrix
\begin{equation}
			\uline{\mathds{M}} = \begin{pmatrix} 
					\ldots & \ldots & \ldots & \ldots & \ldots \\
					\ldots & \mathds{M}_{+} & \mathds{O} & \mathds{O} &\ldots \\
					\ldots &\mathds{O} & \mathds{M} & \mathds{O} &\ldots \\
					\ldots &\mathds{O} & \mathds{O} & \mathds{M}_- &\ldots  \\
                               		\ldots &\ldots & \ldots & \ldots & \ldots
	                     \end{pmatrix}
\end{equation}
and tridiagonal block matrix
\begin{equation}
			\uline{\mathds{L}} = \begin{pmatrix} 
					\ldots & \ldots & \ldots & \ldots & \ldots \\
					\frac{\ri \beta}{2} \mathds{M}_+ \mathds{Q}_+ & \mathds{1} & \frac{\ri \beta}{2} \mathds{M}_+ \mathds{Q}_- & \mathds{O} & \ldots \\
					\ldots &\frac{\ri \beta}{2} \mathds{M} \mathds{Q}_+ & \mathds{1} & \frac{\ri \beta}{2} \mathds{M} \mathds{Q}_- & \ldots \\
					\ldots &\mathds{O} & \frac{\ri \beta}{2} \mathds{M}_- \mathds{Q}_+ & \mathds{1} &\frac{\ri \beta}{2} \mathds{M}_- \mathds{Q}_- \\
                               		\ldots &\ldots & \ldots & \ldots & \ldots
	                     \end{pmatrix}
\end{equation}
we can rewrite the coupled qLE~(\ref{Eq:coupledLangevin}) as a matrix equation
\begin{equation}
	\uline{\mathds{L}} \uline{\boldsymbol{\psi}} = \uline{\mathds{M}} \uline{\mathbf{F}}.
\end{equation}
Hence
\begin{equation}
	\uline{\boldsymbol{\psi}} = \uline{\mathds{L}}^{-1} \uline{\mathds{M}} \uline{\mathbf{F}}.
	\label{Eq:PerturbationSeries}
\end{equation}
By considering only block vectors $\uline{\boldsymbol{\psi}}$ of $2n + 1$ vectors $\boldsymbol{\psi}$ with the corresponding block matrices of size $(2n + 1)\times(2n + 1)$ submatrices we obtain the perturbation results up to order $n$. Note, that the full size of the block vectors and matrices is $N (2n + 1)$ and $N^2 (2n + 1)^2$, resp. 

To evaluate these spectra in our general formalism, we start with Eq.~(\ref{Eq:PerturbationSeries})
and introduce the block matrices $\uline{\mathds{Y}}_1 = {\rm diag}(1,0,0,0,1,0,0,0, \ldots)$, $\uline{\mathds{Y}}_2 = {\rm diag}(0,1,0,0,0,1,0,0, \ldots)$ etc.\ so that $\sum_k \uline{\mathds{Y}}_k = \uline{\mathds{1}}$. These matrices allow us to split the contributions from all baths $k$ so that
\begin{equation}
	\uline{\boldsymbol{\psi}} = \sum_{k = 1}^N \uline{\mathds{L}}^{-1} \uline{\mathds{M}} \mathds{Y}_k \uline{\mathbf{F}}.
\end{equation}
To evaluate products we use the fluctuation-dissipation theorem in the form 
\begin{equation}
  \langle F_{k}^\dagger (\omega +l\Omega) F_{k'}(\omega' + l'\Omega) \rangle = \delta_{k,k'} \delta_{l,l'} 2 \pi \delta(\omega - \omega') \langle F_k^\dagger F_k \rangle_\omega,
\end{equation}
where $\langle F_k^\dagger F_k \rangle_\omega  = 2 \kappa_k n_k$. Here in agreement with the treatment in the qME approach we are assuming that $n_{k}$ is constant as demanded by the assumption of white noise. This assumption is justified for $\beta \ll \omega_k$ and $\Omega \ll \kb T/\hbar$. Then we have 
\begin{equation}
	\langle	\uline{\boldsymbol{\psi}}_\alpha^\dagger \uline{\boldsymbol{\psi}}_\epsilon \rangle_\omega = \sum_{k = 1}^N 2 \kappa_k n_k \bigl( \uline{\mathds{L}}^{-1} \uline{\mathds{M}} \uline{\mathds{Y}}_k  \uline{\mathds{M}}^\dagger {\uline{\mathds{L}}^{-1}}^\dagger \bigr)_{\epsilon,\alpha} 
	\label{Eq:FullSpectrum}
\end{equation}
using the properties $\uline{\mathds{Y}}_{k}^\dagger = \uline{\mathds{Y}}_{k}$ and $\uline{\mathds{Y}}_{k} \uline{\mathds{Y}}_{k} = \uline{\mathds{Y}}_{k}$. From this expression we can numerically calculate all spectral correlation function.

The mean heat flux (transferred power over one oscillation period) from oscillator $k$ at temperature $T_k \neq 0 \,{\rm K}$ to an oscillator $l$ at temperature $T_l = 0\,{\rm K}$ is defined by~\cite{LandiEtAl2022}
\begin{equation}
	P_{k \rightarrow l} = \int \frac{\rd \omega}{2 \pi} \hbar \omega_k 2 \kappa_l \langle a^\dagger_l a_l \rangle_\omega
	\label{Eq:PabLangevin}
\end{equation}
where $\langle a^\dagger_l a_l \rangle_\omega$ is given by $\langle \uline{\boldsymbol{\psi}}_\alpha^\dagger \uline{\boldsymbol{\psi}}_\epsilon \rangle_\omega$ from Eq.~(\ref{Eq:FullSpectrum}) with $ \epsilon = \alpha = Nn + l$ coming from the term involving $n_k$ due to bath $k$. The total power emitted by the hot oscillator $k$ is given by
\begin{equation}
	P_{k}^{\rm em} = \int \frac{\rd \omega}{2 \pi} \hbar \omega_k 2 \kappa_k (n_k - \langle a^\dagger_k a_k \rangle_\omega ).
	\label{Eq:PabLangevinEmitted}
\end{equation}
Since energy conservation is fulfilled the sum of the power received by all resonators $l$ equals that emitted by resonator $k$ so that we have 
\begin{equation}
       	P_{k}^{\rm em} = \sum_{l \neq k} P_{k \rightarrow l}.
\end{equation}
This relation is very useful for validating the numerical results.

\section{Master equations with synthetic fields}

Now, instead of the qLEs we use the qME~(\ref{Eq:aa})-(\ref{Eq:ab}) with a periodic driving as in Eq.~(\ref{Eq:omegai}). This directly leads to the set of equations
\begin{align}
	\frac{\rd}{\rd t} \langle a^\dagger_k  a_k \rangle &= - \ri \sum_{j, j \neq k} \bigl( g_{kj} \langle a^\dagger_k a_j \rangle - g_{jk} \langle a_k a^\dagger_j \rangle \bigr) \nonumber \\ &\quad - 2 \kappa_k \langle a^\dagger_k a_k \rangle + 2 \kappa_k n_k, \label{Eq:aamod}\\
	\frac{\rd}{\rd t} \langle a^\dagger_k a_l\rangle &= \tilde{\Omega}_{kl} \langle a^\dagger_k a_l\rangle - \ri \sum_{j \neq k; j\neq l} \bigl( g_{lj} \langle a^\dagger_k a_j\rangle - g_{jk} \langle a^\dagger_j a_l \rangle  \bigr) \nonumber \\  
	&\quad - \ri g_{lk} \bigl( \langle a_k^\dagger a_k \rangle -\langle a_l^\dagger a_l \rangle \bigr) \label{Eq:abmod}
\end{align}
with
\begin{equation}
	\begin{split}
		\tilde{\Omega}_{kl} &= +\ri (\omega_k - \omega_l) - \kappa_k - \kappa_l\\
		              &\quad + \ri \beta \bigl[ \cos(\Omega t + \theta_k) - \cos(\Omega t + \theta_l) \bigr].
	\end{split}
\end{equation}

To solve the equations we make the Fourier series ansatz for the expectation values of each observable $O$ such that
\begin{equation}
	\langle O \rangle = \sum_n \re^{-\ri n \Omega t} \langle O \rangle_n.
\end{equation}
Then we note that
\begin{equation}
	\begin{split}
		\sum_n \re^{-\ri n \Omega t} &\langle O \rangle_n  \bigl( \cos(\Omega t + \theta_k) - \cos(\Omega t + \theta_l) \bigr) \\
		       &=  \sum_n \re^{-\ri n \Omega t} \biggl[ \frac{\eta_{kl}}{2} \langle O \rangle_{n + 1} +  \frac{\eta^*_{kl}}{2} \langle O \rangle_{n - 1} \biggr]
	\end{split}
\end{equation}
with 
\begin{equation}
  \eta_{kl} = (m_k \re^{\ri \theta_k} - m_l \re^{\ri \theta_l }).
\end{equation}
Inserting this ansatz in the set of Eqs.~(\ref{Eq:aamod})-(\ref{Eq:abmod}) gives the following set of equations for the Fourier components
\begin{align}
	\bigl[ - \ri n \Omega + 2 \kappa_k \bigr] \langle a^\dagger_k a_k \rangle_n &= - \ri \sum_{j, j \neq k} \bigl( g_{kj} \langle a^\dagger_k a_j \rangle_n \nonumber \\ &\qquad\quad - g_{jk} \langle a_k a^\dagger_j \rangle_n \bigr)  \nonumber \\
	            &\quad  + 2 \kappa_k {n}_k \delta_{n0} \label{Eq:Fourier1} \\
	\bigl[ - \ri n \Omega - \Omega_{kl} \bigr] \langle a^\dagger_k a_l \rangle_n &= - \ri \sum_{j \neq k; j\neq l} \bigl( g_{lj} \langle a^\dagger_k a_j\rangle_n \nonumber \\
	           &\qquad \quad- g_{jk} \langle a^\dagger_j a_l \rangle_n  \bigr) \nonumber \\ 
        &\quad - \ri g_{lk} \bigl( \langle a_k^\dagger a_k \rangle_n -\langle a_l^\dagger a_l \rangle_n \bigr) \nonumber \\
	&\quad - \frac{\ri \beta \eta_{kl}}{2} \langle a^\dagger_k a_l\rangle_{n + 1} \nonumber  \\ 
	&\quad - \frac{\ri \beta \eta^*_{kl}}{2} \langle a^\dagger_k a_l\rangle_{n - 1}. 
\end{align}
The set of equations for the Fourier components can again be written in matrix form 
\begin{equation}
	\uuline{\mathds{L}} \uline{\boldsymbol{\psi}} = \uline{\boldsymbol{\kappa}}
\end{equation}
when introducing 
the block vector
\begin{equation}
	\uline{\boldsymbol{\psi}} = (\ldots, \boldsymbol{\psi}_1, \boldsymbol{\psi}_0, \boldsymbol{\psi}_{-1}, \ldots)^t.
\end{equation}
with
\begin{equation}
\begin{split}
	\boldsymbol{\psi}_n &= (\langle a^\dagger_1 a_1 \rangle_n,  \ldots, \langle a^\dagger_N a_N \rangle_n , \\
			    &\quad \langle a^\dagger_1 a_2 \rangle_n , \langle a^\dagger_2 a_1 \rangle_n,  \ldots, \langle a^\dagger_1 a_N \rangle_n,\langle a^\dagger_N a_1 \rangle_n,     \\
                            &\quad \langle a^\dagger_2 a_3 \rangle_n , \langle a^\dagger_3 a_2 \rangle_n,  \ldots, \langle a^\dagger_2 a_N \rangle_n,\langle a^\dagger_N a_2 \rangle_n, \\
			    &\quad \ldots \langle a^\dagger_{N-1} a_N \rangle_n,\langle a^\dagger_N a_{N-1}\rangle)^t
\end{split}
\end{equation}
as well as the block vector
\begin{equation}
	\uline{\boldsymbol{\kappa}} = (\ldots, 0,0, + 2 \kappa_1 {n}_1, \ldots, + 2 \kappa_N {n}_N,0,0, \ldots)^t.
\end{equation}
The block matrix $\uline{\mathds{L}}$ then takes the form of a tri-diagonal block matrix
\begin{equation}
	\uline{\mathds{L}} = 
	\begin{pmatrix}
		\ldots & \ldots & \ldots & \ldots & \ldots \\
	        \ldots & \mathds{M}_1 & \mathds{G}^- & \mathds{O} & \ldots \\
	        \ldots & \mathds{G}^+ & \mathds{M}_0 & \mathds{G}^- & \ldots \\
		\ldots & \mathds{O}& \mathds{G}^+ & \mathds{M}_{-1} & \ldots \\
	\ldots & \ldots & \ldots & \ldots & \ldots
	\end{pmatrix}
\end{equation}
with
\begin{widetext}
\begin{equation}
   \mathds{M}_n = \begin{pmatrix}
	   -\ri n \Omega + 2 \kappa_1 &  0     &\ldots  &0	& +\ri g_{12} 	 & -\ri g_{21} & \ldots & 0 & 0\\
		     0        &-\ri n \Omega + 2 \kappa_2 &\ldots & 0  & -\ri g_{12}    & \ri g_{21} & \ldots & \ldots & \vdots \\
		     \vdots   & \ldots  & \ldots& \ldots &  \ldots & \ldots &  \ldots & \ldots & \vdots \\
		    0 	& 0 & \ldots & -\ri n \Omega + 2 \kappa_N & 0 & 0 & \ldots & - \ri g_{N-1,N} &   \ri g_{N,N - 1} \\
		-\ri g_{21}			   & \ri g_{12} & \ldots&	0 	& -\ri n \Omega -\Omega_{12} & 	0	&\ldots & 0 & 0 \\
		\ri g_{21}			   & - \ri g_{12}	& \ldots & 0		& 				0& -\ri n \Omega  - \Omega_{21}  & \ldots & 0 &0 \\
		\vdots   & \ldots  & \ldots& \ldots &  \ldots & \ldots &  \ldots & \ldots & \vdots \\
		0 & 0 & -\ri g_{3N} & \ri g_{3N} & \ldots & \ldots & \ldots & 0 & -\ri n \Omega -\Omega_{N,N-1}
	        \end{pmatrix} 
\end{equation}
\end{widetext}
and
\begin{equation}
	\mathds{G}^+ = \frac{\ri \beta}{2} {\rm diag} (0, \ldots, 0, \eta_{12}, -\eta_{12}, \ldots, \eta_{N-1,N} , -\eta_{N-1,N}) 
\end{equation}
and $\mathds{G}^-$ defined as the matrix obtained from $\mathds{G}$ when complex conjugating $\eta_{kl}$. The different ``perturbation orders'' $n$ can be obtained by using $2 n + 1$ subblocks in the matrix $\uline{\mathds{L}}$. Note, that even though we use the same notation as in the qLE approach, the used vectors and matrices are different and also have a different dimension. Here the dimension of the block vectors and matrices is $N^2 (2n + 1)$ and $N^4 (2 n + 1)^22$. 

The mean heat flux (transferred power over one oscillation period) from oscillator $k$ at temperature $T_k$ to an oscillator $l$ at temperature $T_l = 0\,{\rm K}$ is defined by~\cite{SABGSA2023}
\begin{equation}
	P_{k \rightarrow l} = \hbar \omega_k 2 \kappa_l \langle a^\dagger_l a_l \rangle_0
	\label{Eq:PabMaster}
\end{equation}
taking $n_i = 0$ for all other resonators. Again the total emitted mean power by particle $k$ is given by
\begin{equation}
	P_{k}^{\rm em} = \hbar \omega_k 2 \kappa_k (n_k - \langle a^\dagger_k a_k \rangle_0)
\end{equation}
and we have energy conservation, i.e.\ $P_{k}^{\rm em} = \sum_{l \neq k} P_{k \rightarrow l}$. The advantage of the qME approach is that, differently from the qLE \eqref{Eq:PabLangevinEmitted}-\eqref{Eq:PabLangevin}, a frequency integration is not necessary. On the other hand, the size of the matrices for a given perturbation order is much larger than for the qLE approach.

\section{Fours resonators case: nonreciprocal heat flux with synthetic fields}

\begin{figure}
  \includegraphics[angle=0,scale=0.25]{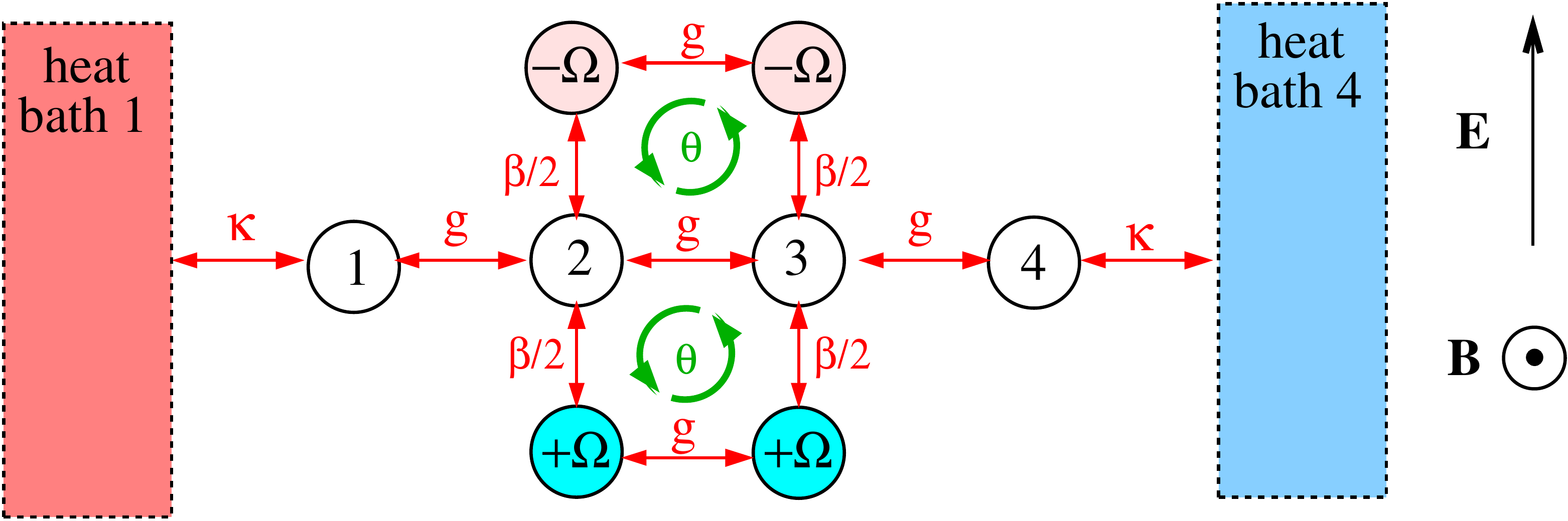}
	\caption{\label{Fig:4Oscillators} Sketch of a chain of four resonators $1$, $2$, $3$, $4$ with equal nearest-neighbor couplings $g$ and resonance frequencies $\omega_0$. The oscillators in the middle are modulated with a modulation strength $\beta$ a relative phase shift $\theta$ resulting in synthetic electric and magnetic fields.}
\end{figure}

We consider here the heat flux in a chain of four resonators as depicted in Fig.~\ref{Fig:4Oscillators}. We assume that all resonators are identical and we further assume reciprocal nearest-neighbour coupling with identical coupling strength $g$ so that the non-zero coupling constants are $g_{12} = g_{21} = g_{32} = g_{23} = g_{34} = g_{43} = g$. The resonance frequencies $\omega_1$ and $\omega_4$ of the resonator $1$ and $4$ are fixed to $\omega_0$, whereas the resonance frequencies of the resonators in the middle are modulated as
\begin{align}
	\omega_2 &= \omega_0 + \beta \cos(\Omega t), \\
	\omega_3 &= \omega_0 + \beta \cos(\Omega t + \theta).
\end{align}
In this configuration, we first determine the power $P_{14}$ transferred from resonator $1$ to resonator $4$ with $T_1 = 300\,{\rm K}$ and $T_{2} = T_{3} = T_{4} = 0\,{\rm K}$. Then we compare with the heat flow in backward direction by calculating the power $P_{41}$ transferred from resonator $4$ to resonator $1$ with $T_4 = 300\,{\rm K}$ and $T_{1} = T_{2} = T_{3} = 0\,{\rm K}$. Hence, only the first and the last resonator are in our configuration coupled to a heat bath. Therefore the modulation frequency $\Omega$ and the modulation strength $\beta$ are here in principle not limited by the constraint due to the white noise assumption because the two resonators in the middle have zero temperature. Nonetheless, we will restrict ourselves to values which fulfill the above criteria for the white noise approximation. For our numerical calculations we use $\omega_0 = 1.69\times10^{14}\,{\rm rad/s}$ and $\kappa = 0.013\omega_0$  which are the the values taken from those for a graphene flake with $E_F = 0.4\,{\rm eV}$ from Ref.~\cite{Graphene}. The coupling constant $g$ is determined by the near-field heat flux value which depends on the relative distance between the graphene flakes. For a distance $d = 100\,{\rm nm}$ between two graphene flakes a fitting of the resonator model with the results from fluctuating electrodynamics~\cite{SABGSA2023} gives $g = 0.011\kappa$. Hence, we are in the weak coupling regime. 

\begin{figure}
  \includegraphics[angle=0,scale=0.7]{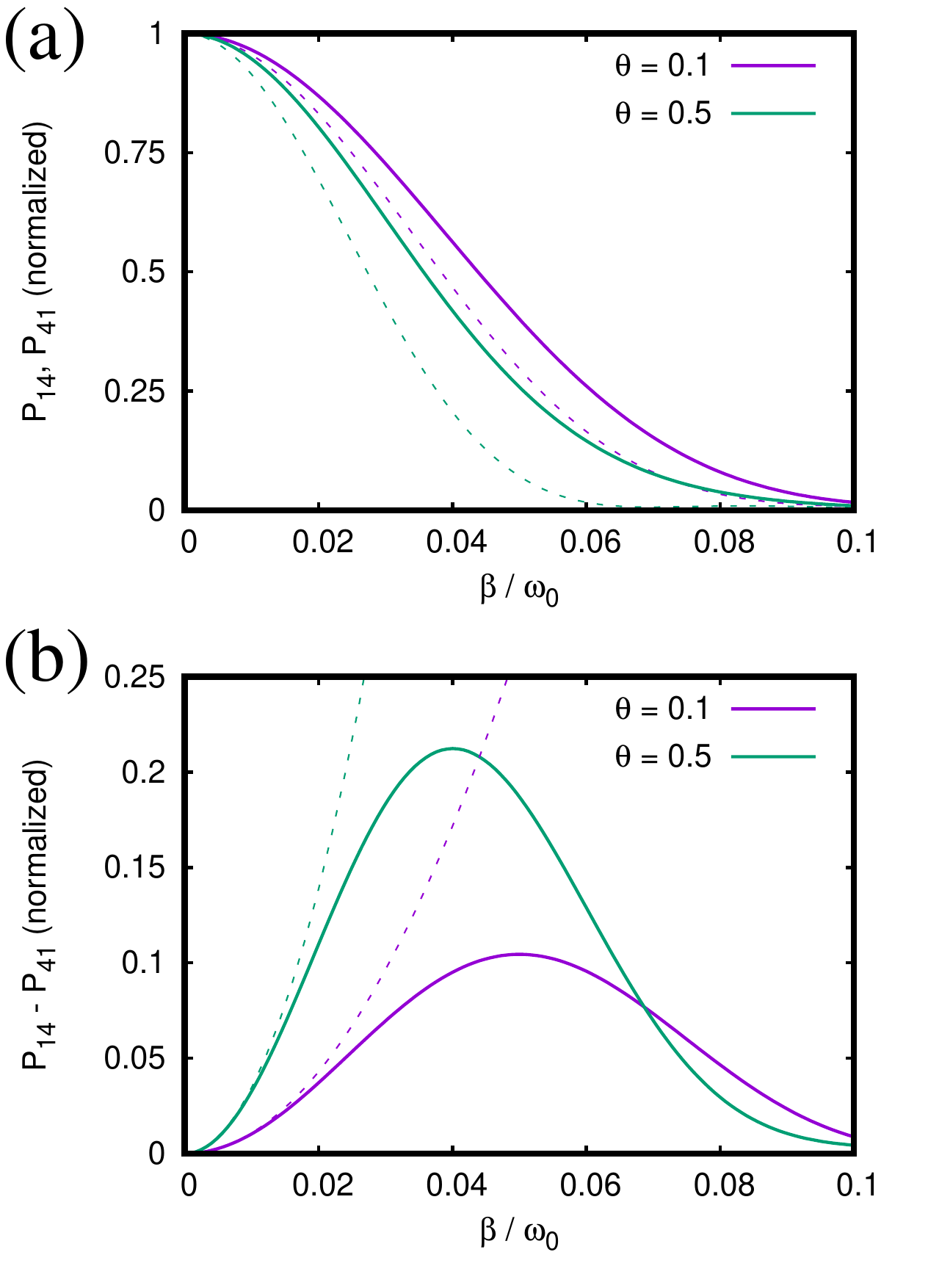}
	\caption{(a) $P_{14}$ (solid line) and $P_{41}$ (dashed line) from the qME approach  (\ref{Eq:PabMaster}) at perturbation order $n = 15$ normalized to the value $P_{14}(\beta = 0) = P_{41}(\beta = 0) = 5.88\times 10^{-22}\,{\rm W}$ for $g = 0.011\kappa$ and $\Omega = 0.05\omega_0$ for $\theta = 0.1 \pi$ and $\theta = 0.5\pi$. The filled and open symbols are the results for $P_{14}$ and $P_{41}$ resulting from integration of spectra as in Fig.~\ref{Fig:Spectra} from qLE approach according to Eq.~(\ref{Eq:PabLangevin}) at perturbation order $n = 10$. (b) Comparison of exact numerical results (solid lines) for the difference $P_{14} - P_{41}$ normalized to $P_{14}(\beta = 0) = P_{41}(\beta = 0) = 5.88\times 10^{-22}\,{\rm W}$ with the corresponding power difference from the approximate expression (dashed lines) from Eq.~(\ref{Eq:ApproxPT}). \label{Fig:PabLangevin}}
\end{figure}

In Fig.~\ref{Fig:PabLangevin}(a) we show the results for the transferred power as function of the modulation strength $\beta$ and for two different values of $\theta$. We also show the numerical results obtained with the qME method with Eq.~(\ref{Eq:PabMaster}) and the qLE approach with Eq.~(\ref{Eq:PabLangevin}). First of all, we can see that both methods provide the same values for the exchanged power. Furthermore, it can be seen that the heat flux is clearly nonreciprocal in contrast to the case of two resonators or two graphene flakes where the heat flux is reciprocal despite the nonreciprocal spectra~\cite{SABGSA2023}.

\begin{figure}
  \includegraphics[angle=0,scale=0.7]{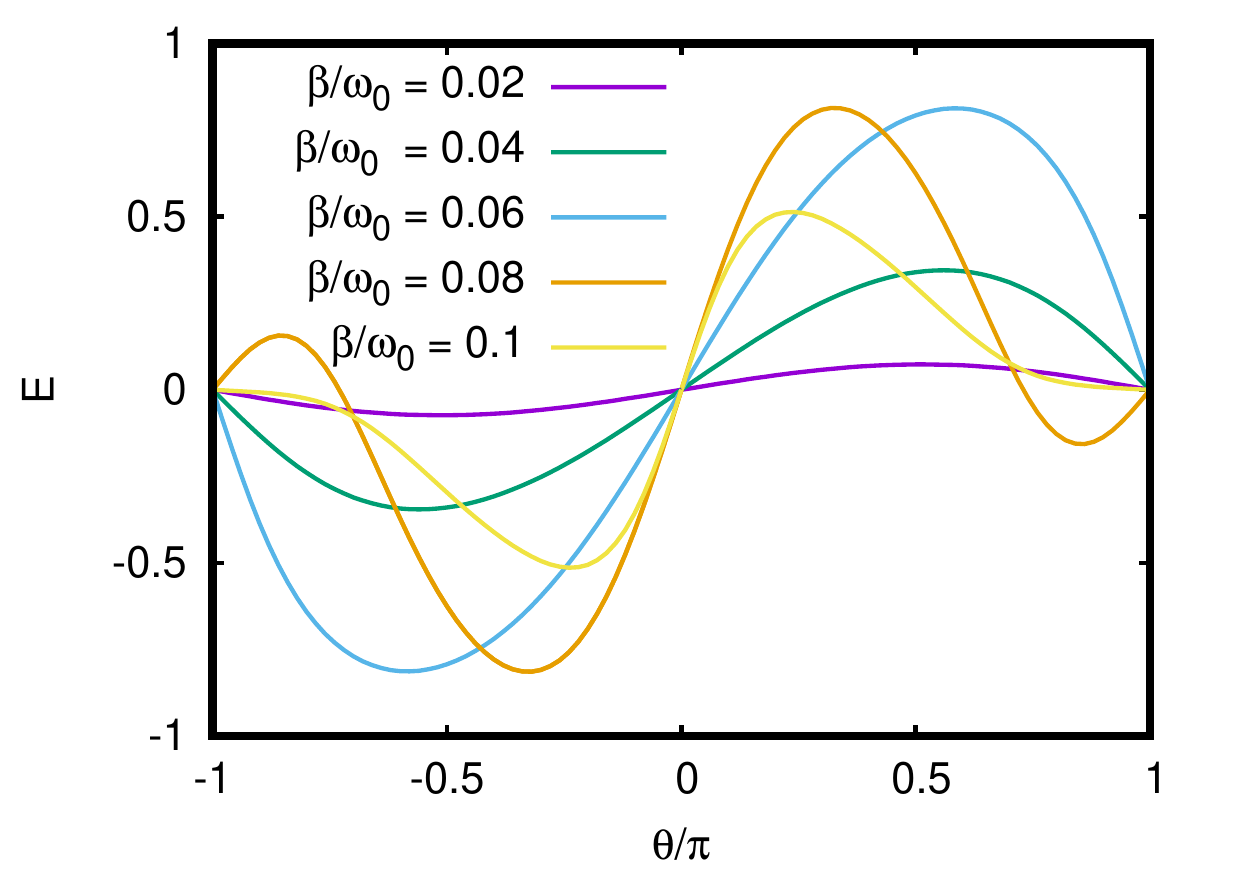}
	\caption{\label{Fig:Theta} We show the relative power transmission $E$ defined in Eq.~(\ref{Eq:relPower}) as function of the dephasing $\theta$ for $\Omega = 0.05\omega_0$ and different values of modulation strength $\beta$ using the qME approach in order $n = 15$.}
\end{figure}

As detailed in Ref.~\cite{PetersonEtAl2019}, for instance, the nonreciprocity in transmission as sketched in Fig.~\ref{Fig:4Oscillators} can be understood in second-order perturbation theory as an interference of different transmission paths. The energy at $\omega_0$  provided by resonator $1$ can go through the chain in second order via the upper and lower sideband at $\omega_0 \pm \Omega$ by two ''scattering events'' $\omega_0 \rightarrow \omega_0 + \Omega$ and  $\omega_0  + \Omega \rightarrow \omega_0$ or  $\omega_0 \rightarrow \omega_0 - \Omega$ and  $\omega_0  - \Omega \rightarrow \omega_0$ as sketched in Fig.~\ref{Fig:4Oscillators}. Due to the presence of the synthetic magnetic field a phase is picked up in this process which is not the same in forward transmission from resonator $1 \rightarrow 4$ and backward transmission from resonator $4 \rightarrow 1$. This symmetry breaking of the synthetic magnetic field can be directly understood from Eq.~(\ref{Eq:SyntheticLagnevin}) which shows that upward and downward transitions in the Floquet side-bands is connected with picking up a positive or negative phase. Hence the forward and backward transmission along the upper or lower sidebands results in different phase factors.
For a plane wave with frequency $\omega$ being transmitted through the coupled resonators $2$ and $3$ the difference in the transmission is explicitely given by~\cite{PetersonEtAl2019} 
\begin{equation}
	\tau_{23} - \tau_{32}  = - 2 \ri \frac{\beta^2}{4} \bigl[\tau(\omega + \Omega)) - \tau(\omega - \Omega))\bigr] \sin(\theta)  
	\label{Eq:Transmission}
\end{equation}
where $\tau(\omega)$ is the transmission coefficient without modulation. This transmission coefficient shows that there is a nonreciprocal transmission for any phase difference $\theta \neq m \pi$ with integer $m$. From this expression it can be expected that at least in second-order perturbation theory, i.e.\ when $\beta$ is sufficiently small, the largest difference can be expected for $\theta = \pi/2$. For the four resonator configuration depicted in Fig.~\ref{Fig:4Oscillators} a similar expression can be derived using a second-order perturbation theory for the qME approach as detailed in appendix~\ref{Appendix}. In the weak coupling limit $g \ll \kappa$ we find for the difference of heat flux in forward and backward direction 
\begin{equation}
\begin{split}
	\frac{P_{14} - P_{41}}{\hbar \omega_0 n g} &=  \beta^2 \frac{g^5}{\kappa^5}  
	\biggl[\frac{7}{8}\frac{\Im(A^2)}{|A|^4}  +  \frac{\kappa \Im(A^3)}{|A|^6} \\
	        &\qquad-\frac{\kappa^3\Im(A^5)}{|A|^{10}}\biggr]\sin(\theta) 
\end{split}
\label{Eq:ApproxPT}
\end{equation}
where $A = 2 \kappa - \ri \Omega$ and $n \equiv n_1 = n_4$ is the mean occupation number of the resonators $1$ in forward or resonator $4$ in backward direction. In Fig.~\ref{Fig:PabLangevin}(b) we compare its predictions with the numerical exact results from Fig.~\ref{Fig:PabLangevin}(a) clearly showing its validity in the small $\beta$ limit. This expression has a similar structure as Eq.~(\ref{Eq:Transmission}) indicating the same dependence on $\theta$ in the limit of small driving amplitudes $\beta$.
To see this effect, we show in Fig.~\ref{Fig:Theta} relative power transmission 
\begin{equation}
   E \equiv \frac{P_{14} - P_{41}}{P_{14} + P_{41}}.
	\label{Eq:relPower}
\end{equation}
It can be seen that indeed for $\beta < 0.05\omega_0$ the maximum difference in forward and backward heat flow happens at $\theta = \pm \pi/2$. For larger modulation strengths higher order effects play a role so that this maximum shifts to slightly larger or smaller values of the dephasing.

\begin{figure}
  \includegraphics[angle=0,scale=0.7]{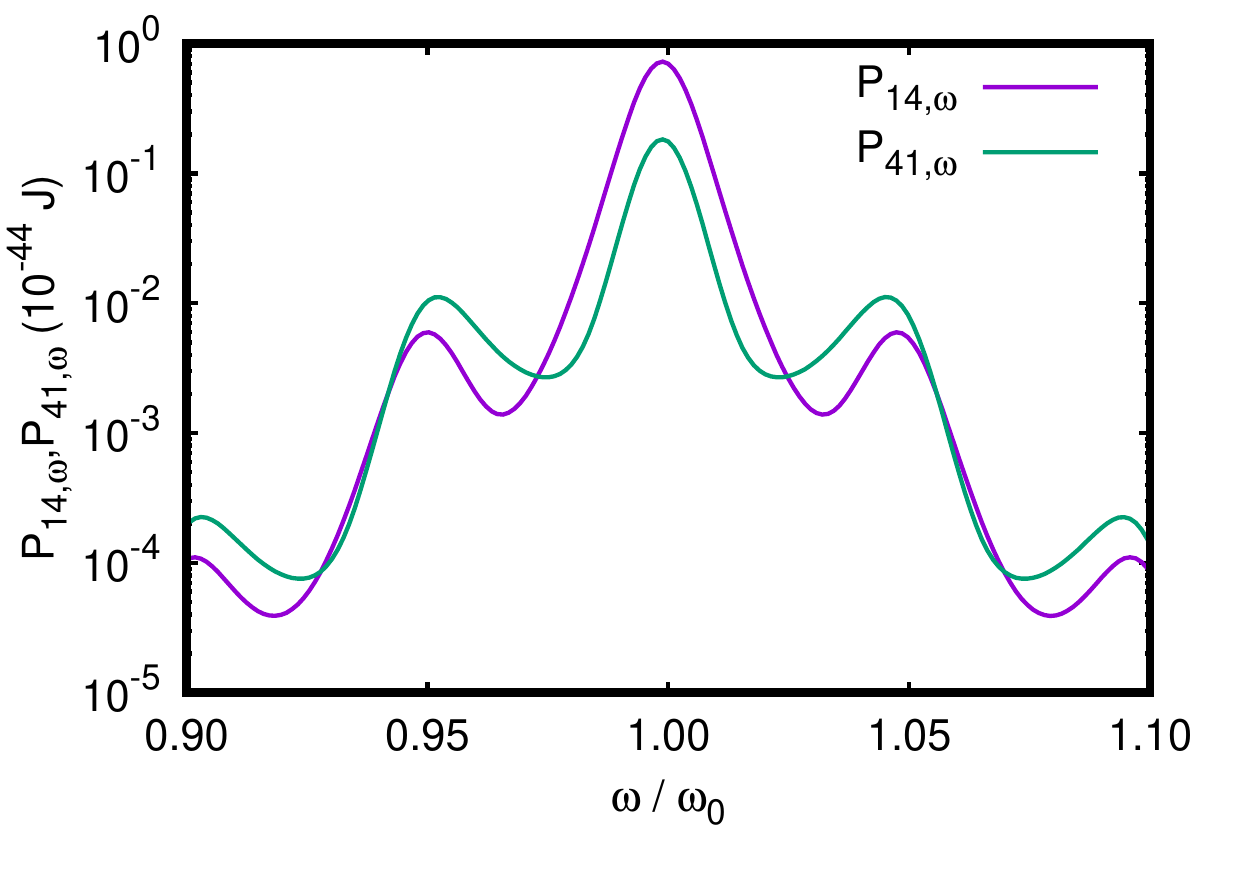}
	\caption{\label{Fig:Spectra} Spectra for mean power $P_{14,\omega} = 2 \kappa \hbar \omega_0 \langle a_4^\dagger a_4 \rangle_\omega$ for the forward heat flow and  $P_{41,\omega} = 2 \kappa \hbar \omega_0 \langle a_1^\dagger a_1 \rangle_\omega$ for the backward heat flow calculated from the spectra for the mean occupations numbers $\langle a_4^\dagger a_4 \rangle_\omega$ and $\langle a_1^\dagger a_1 \rangle_\omega$ in Eq.~(\ref{Eq:FullSpectrum}). The modulation parameters are $\Omega = 0.05\omega_0$, $\beta = 0.05\omega_0$, and $\theta = \pi/2$ and we use perturbation order of $n = 10$.}
\end{figure}

Finally, in Fig.~\ref{Fig:Spectra} the spectra of the power $P_{14,\omega}$ and $P_{41,\omega}$ obtained with the qLE approach in the forward and backward direction are shown using $\Omega = 0.05\omega_0$, $\beta = 0.05\omega_0$, and $\theta = \pi/2$. It can be seen that the spectra for the heat flow in forward and backward dirction are not the same as also found for two graphene flakes only~\cite{SABGSA2023}. Furthermore, it can be seen that the side-band contribution is very small so that the main nonreciprocity stems from frequencies around the resonance $\omega_0$. Integrating these spectra according to Eq.~(\ref{Eq:PabLangevin}) gives the full transferred power for the forward and backward direction shown in Fig.~\ref{Fig:PabLangevin}(a). 

Hence, in our four resonator system we clearly find a nonreciprocal heat flow due to electric and magnetic synthetic fields. Even though our example might be difficult to realize in practice, it clearly shows that synthetic electric and magnetic fields can generate a nonreciprocal heat flux. We emphasize that this result is not limited to near-field heat transfer between graphene flakes but it is generally valid for any configuration and any heat transfer channel which can be described by four coupled resonators with synthetic fields.

\section{Conclusion}

To summarize, based on the local qME we have introduced a formalism for a qLE and qME approach for $N$ coupled resonators with electric and magnetic synthetic fields. The qLE approach is the natural choice when heat flux spectra are studied whereas for the heat flow the qME approach is a better choice, because it is faster. As a very important example, we have used both approaches to show for a system of four linearly coupled resonators that the heat flow is nonreciprocal when electric and magnetic synthetic fields are present. We have also verified numerically that both approach give the same values for the heat flux. Even though for the numerical evaluation we have considered the near-field heat transfer in a system of four coupled graphene flakes our findings are very general and applicable to any system and any heat flux channel which can be described by coupled resonators. Hence, our formalism provides the fundament for further studies on the heat flux and other physical effects in coupled many resonator systems with synthetic fields.

\acknowledgments

S.-A.\ B.\ acknowledges support from Heisenberg Programme of the Deutsche Forschungsgemeinschaft (DFG, German Research Foundation) under the project No.\ 461632548. S.-A.\ B.\ and P. R.-L.\ thank the University of Montpellier and the group Theory of Light-Matter and Quantum Phenomena of the Laboratoire Charles Coulomb for hospitality during his stay in Montpellier where parts of this work has been done. S.-A.~B., P.~R.-L., and M.~A.\ acknowledge QuantUM program of the University of Montpellier. G.S.~A.\ thanks the kind support of The Air Force Office of Scientific Research [AFOSR award no. FA9550-20-1-0366] and The Robert A. Welch Foundation [grant no. A-1943]. P.~R.-L. acknowledges support from AYUDA PUENTE 2022, URJC. 

\appendix

\section{Perturbation theory for qME approach}\label{Appendix}

In this section we derive the second order expression in Eq.~(\ref{Eq:ApproxPT}). To this end, we start with Fourier equations for the qME in (41) taking terms with $n = 0,1,-1$. Then we have
\begin{align}
	\mathds{M}_0 \boldsymbol{\psi}_0 &= \boldsymbol{\kappa} - \mathds{G}^+ \boldsymbol{\psi}_{+1} - \mathds{G}^- \boldsymbol{\psi}_{-1}, \\
	 \mathds{M}_{+1} \boldsymbol{\psi}_{+1} &=  - \mathds{G}^+ \boldsymbol{\psi}_2 - \mathds{G}^- \boldsymbol{\psi}_{0}, \\
	  \mathds{M}_{-1} \boldsymbol{\psi}_{-1} &=  -  \mathds{G}^+ \boldsymbol{\psi}_0 - \mathds{G}^- \boldsymbol{\psi}_{-2}.
\end{align}
By inserting the expressions for $\boldsymbol{\psi}_{+1/-1}$ into the equation for $\boldsymbol{\psi}_0$ and neclecting terms from $|n| \geq 2$ we arrive at
\begin{equation}
	\mathds{N} \boldsymbol{\psi}_0 =  \boldsymbol{\kappa} \quad \Rightarrow \quad \boldsymbol{\psi}_0 = \mathds{N}^{-1} \boldsymbol{\kappa} 
\end{equation}
with
\begin{equation}
  	\mathds{N} = \bigl[ \mathds{M}_0 -  \mathds{G}^+ \mathds{M}_{+1}^{-1}  \mathds{G}^- -  \mathds{G}^- \mathds{M}_{-1}^{-1} \mathds{G}^+ \bigr]. 
\end{equation}
By defining 
\begin{equation}
	\mathds{G}^+ = \frac{\ri \beta}{2}\tilde{\mathds{G}} \quad \text{and} \quad \mathds{G}^- = \frac{\ri \beta}{2}\tilde{\mathds{G}}^* 
\end{equation}
with $\tilde{G} = {\rm diag} (0,\ldots,0, \eta_{12}, -\eta_{21}, \ldots, \eta_{N-1,N}, -\eta_{N-1,N})$
we have 
\begin{equation}
	\mathds{N} = \biggl[ \mathds{M}_0 + \frac{\beta^2}{4} \bigl( \tilde{\mathds{G}} \mathds{M}_{+1}^{-1}  \tilde{\mathds{G}}^* +  \tilde{\mathds{G}}^* \mathds{M}_{-1}^{-1} \tilde{\mathds{G}} \bigr) \biggr]. 
\end{equation}
From this expressions it becomes more obvious that the first non-vanishing contributions to the zeroth order are stemming from the second-order terms, i.e.\ there is no contribution linear in $\beta$.

\begin{figure}
  \includegraphics[angle=0,scale=0.7]{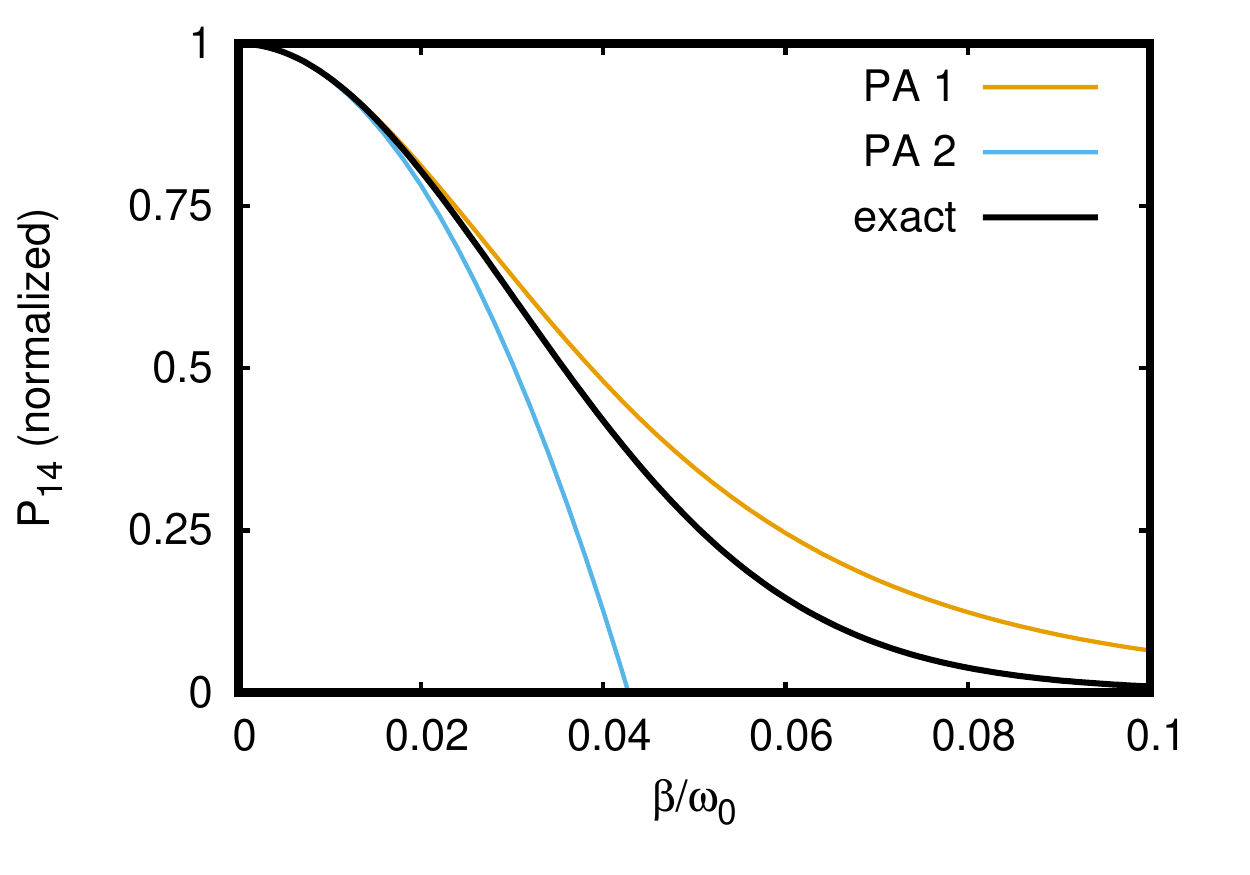}
	\caption{\label{Fig:Pert} Comparison of exact numerical results for $P_{14}$ (black lines) with second order perturpation approach from Eq.~(A8) (PA 1) and with those from Eq.~(A9) (PA 2) using the same parameters as in Fig.~\ref{Fig:PabLangevin}(a) and $\theta = \pi/2$. The approximations for $P_{41}$ are similar (not shown).}
\end{figure}

For the tight-binding model of the four identical resonators the involved vectors have 16 components and the matrices have a size of $16\times16$. By definition of $ \boldsymbol{\psi}_0$ we are interested in the terms $\mathds{N}^{-1}_{14}$ and  $\mathds{N}^{-1}_{41}$ which are determining the transferred power $P_{4 \rightarrow 1}$ and $P_{1 \rightarrow 4}$. Obviously, there can only be a non-reciprocity if $\mathds{N}^{-1} \neq {\mathds{N}^{-1}}^t$. From the equation for $\mathds{N}$ it can be seen that due to the phase terms $\tilde{\mathds{G}}$ and $\tilde{\mathds{G}}^*$ in the second-order contribution, in general, we have $\mathds{N} \neq \mathds{N}^t$ so that also $P_{4 \rightarrow 1} \neq P_{1 \rightarrow 4}$ in general. Hence, the synthetic magnetic field results in an asymmetry for $\mathds{N}$ and hence for $\mathds{N}^{-1}$.

For small $\beta$ we can further simplify the inverse of $\mathds{N}$ in Eq.~(A8) 
\begin{equation}
\begin{split}
	\mathds{N}^{-1} &= \biggl[ \mathds{M}_0 + \frac{\beta^2}{4} \bigl( \tilde{\mathds{G}} \mathds{M}_{+1}^{-1}  \tilde{\mathds{G}}^* +  \tilde{\mathds{G}}^* \mathds{M}_{-1}^{-1} \tilde{\mathds{G}} \bigr) \biggr]^{-1} \\
			&=  \biggl[ \mathds{1} + \frac{\beta^2}{4}  \mathds{M}_0^{-1} \bigl( \tilde{\mathds{G}} \mathds{M}_{+1}^{-1}  \tilde{\mathds{G}}^* +  \tilde{\mathds{G}}^* \mathds{M}_{-1}^{-1} \tilde{\mathds{G}} \bigr) \biggr]^{-1} \mathds{M}_0^{-1}\\
			&\approx \biggl[ \mathds{1} - \frac{\beta^2}{4}  \mathds{M}_0^{-1} \bigl( \tilde{\mathds{G}} \mathds{M}_{+1}^{-1}  \tilde{\mathds{G}}^* +  \tilde{\mathds{G}}^* \mathds{M}_{-1}^{-1} \tilde{\mathds{G}} \bigr) \biggr] \mathds{M}_0^{-1}.
\end{split}
\end{equation}
In Fig.~\ref{Fig:Pert} we show a comparison of the second-order results using Eq.~(A8) and Eq.~(A9) with the numrical exact results. As expected the second-order expansion is only reliable for small enough values of $\beta$ and the perturbation expression in Eq.~(A8) is valid for a larger range than the perturbative expression in Eq.~(A9).

Now, we want to derive an analytical expression for the heat flux difference. Note, that the heat flux difference for the forward and backward case is in our example given by
\begin{equation}
	P_{14} - P_{41} = 4 \hbar \omega_0 n \kappa^2  \Delta N_{14} 
	\label{Eq:AppendixDiff}
\end{equation}
where $\Delta N_{14} = \mathds{N}_{14}^{-1} - \mathds{N}_{41}^{-1}$ and $n \equiv n_1 = n_4$. That means we can focus on $\Delta N_{14}$ and add the prefactors later. Starting with the approximate expression in Eq.~(A9) and making a Taylor expansion for $g \ll \kappa$ we obtain with Mathematica for $\Delta N_{14}$ the relatively long expression
\begin{widetext}
\begin{equation}
\begin{split}
	\Delta N_{14} \approx \frac{\beta^2 g^2}{8 |A_1|^6} \frac{g^4}{\kappa^4}&
	\biggl[\frac{ |A_{1}|^2 \Im(A_1^2)}{A_0^3} \biggl(4 \bigl[ \Im( \eta_{13} \eta_{12}^*) + \Im(\eta_{34} \eta_{24}^*)] + 3 \bigl[ \Im(\eta_{23} \eta_{13}^*) + \Im(\eta_{24} \eta_{23}^*) \bigr]  + \Im(\eta_{14} \eta_{13}^*) + \Im(\eta_{24} \eta_{14}^*)\biggr) \\
	 &+ \frac{\Im(A_1^3)}{A_0^2} \biggl(\Im(\eta_{14} \eta_{12}^*) + 2 \Im(\eta_{24} \eta_{13}^*)+ \Im(\eta_{34} \eta_{14}^*) - 3 \Im(\eta_{12} \eta_{23}^*) -3 \Im(\eta_{23} \eta_{34}^*) \biggr) \\
	 &+2 \frac{\Im(A_1^4)}{|A_{1}|^2 A_0} \biggl(\Im(\eta_{24} \eta_{12}^*) + \Im(\eta_{34} \eta_{13}^*) \biggr) -\frac{2 \Im(A_{1}^5)}{|A_{1}|^4} \Im(\eta_{12} \eta_{34}^*)\biggr] 
\end{split}
\end{equation}
\end{widetext}
where we have introduced $A_n = 2 \kappa - \ri n \Omega$. From this expression it can be seen that only for complex $\eta_{ij}$ there is a non-reciprocity. It can be further observed that there seem to be plenty of combinations which give a non-reciprocal heat flux. In our four oscillator example the resonator $3$ is the only one with a nonzero phase $\theta \equiv \theta_3 \neq 0$ and resonator $1$ and $4$ are not modulated at all so that $\eta_{12} = -1$, $\eta_{14} = 0$, $\eta_{24} = 1$ and $\eta_{34} = \re^{\ri \theta} = - \eta_{13}$ and $\eta_{23} = 1 - \re^{\ri \theta}$. With these specific values we get 
\begin{equation}
\begin{split}
	\Delta N_{14} \approx \frac{\beta^2 g^6}{4\kappa^4}  \sin(\theta)&
	\biggl[\frac{7\Im(A_1^2)}{|A_1|^4 A_0^3}  + \frac{4\Im(A_1^3)}{|A_1|^6A_0^2} -\frac{\Im(A_{1}^5)}{|A_{1}|^{10}}\biggr]. 
\end{split}
\label{Eq:Approx}
\end{equation}
By adding the correspondig factors as defined in Eq. (\ref{Eq:AppendixDiff}) and realizing that $A_0 = 2 \kappa$ we obtain the approximative analytical expression for the heat flux difference in Eq.~(\ref{Eq:ApproxPT}).

\end{document}